# Planar Thermal Hall Effects in Kitaev Spin Liquid Candidate Na$_2$Co$_2$TeO$_6$

Authors and Affiliations:

Hikaru Takeda[1], Jiancong Mai[1], Masatoshi Akazawa[1], Kyo Tamura[1], Jian Yan[1], Kalimuthu Moovendaran[2], Kalaivanan Raju[2], Raman Sankar[2], Kwang-Yong Choi[3], and Minoru Yamashita[1]

1 The Institute for Solid State Physics, University of Tokyo, Kashiwa, 277-8581, Japan
2 Institute of Physics, Academia Sinica, Taipei 11529, Taiwan
3 Department of Physics, Sungkyunkwan University, Suwon 16419, Republic of Korea

Abstract:

We investigate both the longitudinal thermal conductivity ($\kappa_{xx}$) and the planar thermal Hall conductivity ($\kappa_{xy}$) in the Kitaev spin liquid candidate of Co-based honeycomb antiferromagnet Na$_2$Co$_2$TeO$_6$ in a magnetic field ($B$) applied along the $a$ and $a^*$ axes. A finite $\kappa_{xy}$ is resolved for both field directions in the antiferromagnetic (AFM) phase below the Néel temperature of 27 K. The temperature dependence of $\kappa_{xy}/T$ shows the emergence of topological bosonic excitations. In addition, the field dependence of $\kappa_{xy}$ shows sign reversals at the critical fields in the AFM phase, suggesting the changes in the Chern number distribution of the topological magnons. Remarkably, a finite $\kappa_{xy}$ is observed in $B \parallel a^*$ between the first-order transition field in the AFM phase and the saturation field, which is prohibited in a disordered state by the two-fold rotation symmetry around the $a^*$ axis of the honeycomb lattice, showing the presence of a magnetically ordered state that breaks the two-fold rotation symmetry. Our results demonstrate the presence of topological magnons in this compound in the whole field range below the saturation field.

Main texts:

Topology in condensed-matter physics is a powerful concept allowing one to characterize material properties solely by the topological classification without details of materials. For conduction electrons in a metal, the anomalous quantum Hall effect [1] is one of the most celebrated examples of such topological phenomena, in which the topological invariant (called the Chern number) of conduction electrons is responsible for the dissipationless quantized Hall current even in the zero field [2]. In an insulator, topological effects on the heat carriers give rise to thermal Hall effects (THEs) as described by [3,4]

$$\frac{\kappa_{xy}}{T} = \frac{k_B^2}{\hbar} \int \Omega(E) f(E) dE, \tag{1}$$

where $\Omega(E)$ is the energy distribution of the Berry curvature and $f(E)$ is given by the distribution function of the elementary excitations at the energy $E$, demonstrating that one can reveal the topological property of the charge-neutral excitations by the thermal Hall measurements.

For fermions, the Fermi distribution leads to quantized transport dictated by the sum of the Chern numbers of the occupied bands below the Fermi energy. For example, in the quantum spin liquid (QSL) given by the Kitaev Hamiltonian, the nontrivial topology of the Majorana fermions gives rise to the half quantization of the thermal Hall conductivity ($\kappa_{xy}$) [5]. In fact, the half-quantized $\kappa_{xy}/T$ in the Kitaev QSL, which is suggested to be realized in Ru or Ir compounds [6], has been reported in α–RuCl$_3$ by several groups [7–10]. For bosons, on the other hand, their topological transport vanishes in the zero-temperature limit, because the temperature dependence is governed by the Bose distribution function [11]. Such topological THEs of bosons are proposed for topological magnons [12–15], triplons [16], and skyrmions [17,18]. Indeed, the topological THE of magnons is reported in a lattice of magnetic skyrmions lately [19].

Recently, new Kitaev QSL candidates have been put forward for $3d$ compounds [20–22]. The Co-based honeycomb antiferromagnet Na$_2$Co$_2$TeO$_6$ [23] constitutes a prominent candidate, in which the high-spin $d^7$ state of Co$^{2+}$ ions forms two-dimensional honeycomb layers of $J_\text{eff} = 1/2$ Kramers doublets with a sizable Kitaev interaction term estimated from the neutron scattering experiments [24–27]. At zero field, zigzag [28–30] or triple-q [31,32] antiferromagnetic (AFM) order occurs below the Néel temperature $T_\text{N} = 27$ K. Magnetization measurements under in-plane magnetic fields [25,30,33] show multiple phase transitions at three critical fields in the AFM phase, as well as a possible emergence of a quantum disordered state between the AFM phase and the spin polarized phase above the saturation field ($B_\text{sat} \sim 13$ T). The longitudinal thermal conductivity measurements performed under in-plane fields [33] and the thermal Hall conductivity measurements under $B \parallel c$ [34,35] bear a close resemblance with those of α–RuCl$_3$, implying a possible realization of the Kitaev QSL in Na$_2$Co$_2$TeO$_6$ under in-plane field above $\sim 10$ T. However, the half-quantized $\kappa_{xy}/T$ under the in-plane field, the key signature of the Kitaev QSL, has been missing.

In this Letter, we address the planar THE of Na$_2$Co$_2$TeO$_6$ under in-plane magnetic fields. Our salient findings are (1) a finite planar $\kappa_{xy}$ for both $B \parallel a$ and $B \parallel a^*$, showing the emergence of topological charge-neutral excitations in Na$_2$Co$_2$TeO$_6$, (2) the vanishing temperature dependence of $\kappa_{xy}/T$ in the zero-temperature limit, consistent with topological bosonic excitations rather than Majorana fermions, and (3) multiple sign changes of $\kappa_{xy}$ at critical fields in $B \parallel a$, suggesting changes of the Berry curvature of the topological magnons in different magnetic states.

Single crystals of $Na_2Co_2TeO_6$ were synthesized by the self-flux method as described in Ref. [32]. Prior to the thermal-transport measurements, we confirmed the crystallographic directions by x-ray diffraction measurements. Thermal-transport measurements were performed by the steady method as described in Ref. [36] by using a variable temperature insert (VTI) for 2–60 K and a dilution refrigerator (DR) for 0.1–3 K. The heat current $J_Q$ and the magnetic field $B$ were applied parallel to either the $a$ and $a^*$ axes of the sample (see Fig. S1 in Supplementary Materials (SM) for more details). The magnetic susceptibility measurements were performed by using a SQUID magnetometer to check the quality and the crystallographic directions of the sample. The reproducibility was confirmed by repeating the $\kappa_{xy}$ measurements in different single crystals (see Figs. S4 and S5 in SM).

Figure 1 shows the temperature dependence of the longitudinal thermal conductivity ($\kappa_{xx}$) measured under $B \parallel J_Q \parallel a$ ("zigzag" direction, perpendicular to the honeycomb bond, Fig. 1(a)) and $B \parallel J_Q \parallel a^*$ ("armchair" direction, parallel to the honeycomb bond, Fig. 1(b)). At zero field, the temperature dependence of $\kappa_{xx}$ shows a broad peak around 50 K, which is followed by a kink at $T_N$, as clearly seen in the temperature dependence of $\kappa_{xx}/T$ (see Fig. S2 in SM). The Néel transition of the sample is also clearly seen as the peak in the temperature dependence of the magnetic susceptibility (Fig. S3 in SM). The good quality of the sample is also confirmed by the boundary-limited phonons observed $\kappa_{xx}$ at 15 T (Fig. S6 in SM) where the phonon thermal conduction is enhanced by suppressing a magnon-phonon scattering by opening a magnon gap (see section IV in SM). This enhancement of $\kappa_{xx}$ at 15 T (Fig. 1) shows the presence of a strong spin-phonon coupling suppressing $\kappa_{xx}$ at 0 T in this compound. The residual of $\kappa_{xx}/T$ in the zero-temperature limit is vanishingly small for the whole field range we measured (see Fig. S7 in SM), showing the absence of itinerant gapless excitations.

Figure 2 shows the field dependence of $\kappa_{xx}$ at fixed temperatures. The field dependence of $\kappa_{xx}$ is negligible above $T_N$ except for the small increase at the highest field. This increase of $\kappa_{xx}$ could be attributed to an increase of the phonon conduction caused by a field suppression effect on the spin fluctuation. Below $T_N$, $\kappa_{xx}$ starts to depend on the magnetic field with features at critical fields of the AFM phases determined by the previous studies [25,30,33] ($B_{c1}$ and $B_{c2}$ for $B \parallel a$, and $B_{c1}^*$, $B_{c2}^*$, and $B_{c3}^*$ for $B \parallel a^*$), as marked by arrows in Fig. 2. In $B \parallel a$, the field dependence of $dM/dB$ shows a shoulder at $B_{c1} \sim 7.6$ T and a peak at $B_{c2} \sim 9.8$ T [33]. Corresponding to these critical fields, $\kappa_{xx}$ shows the peak at $B_{c1}$ and the dip at $B_{c2}$, as shown in Fig. 2(a). On the other hand, in $B \parallel a^*$, three critical fields (denoted as $B_{c1}^*$, $B_{c2}^*$, and $B_{c3}^*$) are found in the field dependence of $dM/dB$ [33]. As shown in Fig. 2(b), $\kappa_{xx}$ shows a sharp jump at $B_{c1}^* \sim 6$ T with a large magnetic hysteresis, demonstrating the first-order transition at $B_{c1}^*$. This first-order transition is also observed as the jump in the magnetization of our sample in $B \parallel a^*$ (Fig. S3(b) in SM) [25,30,33]. In contrast to $\kappa_{xx}$ in $B \parallel a$, only small humps are observed at $B_{c2}^*$ and $B_{c3}^*$ in $B \parallel a^*$. As $\kappa_{xx}$ of a magnetic insulator is given by the sum of the phonon contribution $\kappa_{xx}^{\text{ph}}$ and the magnon contribution $\kappa_{xx}^{\text{mag}}$, the field dependence of $\kappa_{xx}$ of this compound can be understood in terms of the field effect of the magnon gap on $\kappa_{xx}^{\text{ph}}$ and $\kappa_{xx}^{\text{mag}}$ (see section VI in SM).

Our main finding is the finite planar $\kappa_{xy}$ in the AFM phase for both $B \parallel a$ ($\kappa_{xy}^a$, Fig. 3(a)) and $B \parallel a^*$ ($\kappa_{xy}^{a^*}$, Fig. 3(b)). We averaged the data of $\kappa_{xy}^a/T$ obtained in the field-up and field-down procedures except for that measured in the DR ($T < 2$ K). On the other hand, we present only the field-up data of $\kappa_{xy}^{a^*}/T$ to avoid mixing the hysteretic field dependence at $B_{c1}^*$ (see Fig. S8 in SM for all the data in each measurement). As shown in Fig. 3, whereas $\kappa_{xy}/T$ is virtually absent above $T_N$, a finite $\kappa_{xy}/T$ is observed in the AFM phase below $T_N$ in both field directions. In $B \parallel a$, $\kappa_{xy}^a/T$ gradually increases up to $B_{c1}$, which is followed by a sign change to negative $\kappa_{xy}^a$ at $B_{c1}$, and another sign change to positive $\kappa_{xy}^a$ at $B_{c2}$. In $B \parallel a^*$, $\kappa_{xy}^{a^*}$ shows a sharp negative peak at the first-order transition at $B_{c1}^*$, which is followed by a constant negative $\kappa_{xy}^{a^*}$ up to the highest fields. We note that the sharp negative anomaly of $\kappa_{xy}^{a^*}$ at $B_{c1}^*$ at 2 K is more pronounced in the field-down measurements (see Fig. S8 in SM). We also note that the planar THE observed in Na$_2$Co$_2$TeO$_6$ is completely different from "planar Hall effects" studied in ferromagnets [37] and Weyl semimetals [38] which are *symmetric* with respect to the field direction. In sharp contrast, the planar THE observed in this compound only comes from the *asymmetric* component in the transverse temperature difference (see section I in SM).

To investigate the origin of the planar THE, we turn to the temperature dependence of $\kappa_{xy}/T$ at the selected fields (see arrows in Fig. 5). As shown in Fig. 4, the temperature dependence of $\kappa_{xy}/T$ shows a peak at 4–10 K, which is followed by a decrease of $|\kappa_{xy}/T|$ to zero as $T \to 0$ K for both field directions. Remarkably, we find that, although $\kappa_{xx}/T$ at 10 T features one phonon peak at approximately 30 K, $\kappa_{xx}/T$ at 7, 9, and 11 T shows the second lower-temperature peak at approximately 5 K, which almost coincides with the peak of $\kappa_{xy}/T$ at 7 and 11 T.

Let us first focus on the origin of the planar THE observed in Na$_2$Co$_2$TeO$_6$, which could be caused by Majorana fermions [5], phonons [39], or topological magnons [12–15]. The magnitude of $\kappa_{xy}/T$ of the Majorana fermions in the Kitaev QSL is expected to show the half-quantized value $\kappa_{xy}^{2D}/T = (\pi/12)(k_B^2/\hbar)$, where $k_B$ is the Boltzmann constant, and $\hbar$ the reduced Planck constant. This half-quantized thermal Hall conductance per one layer corresponds to $\kappa_{xy}/T = (\kappa_{xy}^{2D}/d)/T \sim 8.47 \times 10^{-4}$ W K$^{-2}$ m$^{-1}$ in this compound, where $d = 0.559$ nm is the interlayer distance [26].

However, as shown in Fig. 3 and Fig. 4, $\kappa_{xy}/T$ observed in Na$_2$Co$_2$TeO$_6$ remains about one order of magnitude smaller than the half-quantized value. In addition, although $\kappa_{xy}/T$ of fermions should stay constant in the zero-temperature limit, $\kappa_{xy}/T$ of Na$_2$Co$_2$TeO$_6$ approaches zero with lowering temperature (Fig. 4), showing the bosonic nature of the elementary excitations. In α-RuCl$_3$, it is pointed out that the half-quantized $\kappa_{xy}/T$ is suppressed in the low-quality samples with suppressed $\kappa_{xx}^{\mathrm{ph}}$ [8,9]. This is clearly not the case for our sample in which the phonon with a long mean free path (Fig. S6 in SM) is observed. Indeed, the magnitude of $\kappa_{xx}$ of our sample at 15 T (Figs. 1(a) and 1(b)) and that of the sample showing the half-quantized $\kappa_{xy}$ in α-RuCl$_3$ below the Néel temperature [8], in which $\kappa_{xx}^{\mathrm{ph}}$ without the strong magnon-phonon scattering specific to each materials can be observed, are similar in these two materials (approximately 6 W K$^{-1}$ m$^{-1}$ at 5 K, for example). It is also pointed out that a good thermalization between the Majorana fermions and phonons is necessary to observe the half-quantized $\kappa_{xy}/T$ in the Kitaev QSL [40,41]. Although an accurate estimation for the Majorana-phonon coupling in Na$_2$Co$_2$TeO$_6$ is not possible at this moment, this coupling should be related to the spin-phonon coupling that gives rise to the positive magneto-thermal conductivity. As shown in Fig. 2, the thermal conductivity at 5 K increases by a factor of 8 under the in-plane field of 15 T, which is much larger than that in α-RuCl$_3$ at a similar temperature and field [42]. Therefore, it is implausible that the suppressed $\kappa_{xy}/T$ is caused by a much smaller coupling between the Majorana fermions and phonons in Na$_2$Co$_2$TeO$_6$ than that in α-RuCl$_3$. Furthermore, a finite $\kappa_{xy}$ is prohibited in $B \parallel a^*$ by the two-fold rotation symmetry around the $a^*$ axis in a Kitaev-Heisenberg paramagnet [13–15], demonstrating the presence of a long-range ordered state that breaks the two-fold rotation symmetry in $B \parallel a^*$. Therefore, we conclude that the emergence of Majorana fermions is unlikely in Na$_2$Co$_2$TeO$_6$.

As shown in Fig. 4, the planar THE develops below $T_{\mathrm{N}}$, suggesting a dominant contribution from topological magnons in Na$_2$Co$_2$TeO$_6$. The abrupt sign changes of $\kappa_{xy}$ observed at the magnetic phase boundaries in the AFM phase (Fig. 5(b)) also suggest that the planar THE arises from the intrinsic Berry phase effect of the topological magnons, because the sign changes can be most naturally attributed to a redistribution of the Berry curvature of the magnons by the magnetic transitions. In fact, the numerical calculations demonstrate sign changes of $\kappa_{xy}$ of the magnons at the magnetic phase transitions in the Kitaev-Heisenberg model [13,15,43]. On the other hand, it would be unlikely that the abrupt sign changes are caused by a reversal of scattering direction upon collisions of magnons with extrinsic impurities at the magnetic transitions.

Remarkably, a recent report of $\kappa_{xy}$ done in $B \parallel c$ [35] shows a very different temperature dependence from our planar THE. In $B \parallel c$, a large $\kappa_{xy}$ emerges far above $T_{\mathrm{N}}$ with a similar temperature dependence with that of $\kappa_{xx}$. This good scaling between $\kappa_{xy}$ and $\kappa_{xx}$ is regarded as the key feature of the phonon THEs [44,45], suggesting a dominant phonon contribution in $\kappa_{xy}$ under $B \parallel c$ which could be caused by the strong spin-phonon coupling suppressing $\kappa_{xx}$. In sharp contrast, the peak temperature of our planar $\kappa_{xy}/T$ (4–10 K) is far below the phonon peak of $\kappa_{xx}/T$ at about 30 K above $T_\mathrm{N}$ (Fig. 4), supporting that the dominant contribution of the planar THE comes from the topological magnons rather than phonons acting in $B \parallel c$.

We thus conclude that the planar THE observed in Na$_2$Co$_2$TeO$_6$ mainly stems from the topological magnons in long-range ordered phases [12–15], whereas, given the current insufficient knowledge about the details of the magnetic order and the spin Hamiltonian of this compound, it remains as future work to identify the magnon bands that can reproduce the $\kappa_{xy}$ data. It should be noted that phonons may play a certain role in assisting the planar THE through spin-phonon coupling [46–50]. As shown in Fig. 4(b), the peak temperature of $\kappa_{xy}/T$ at 7 and 11 T coincides with the second peak of $\kappa_{xx}/T$ caused by the field enhancing effect on $\kappa_{xx}^{\mathrm{ph}}$ (section VI in SM). This coincidence may suggest that $\kappa_{xy}$ is enhanced thorough a spin-phonon coupling to the magnetic moment that develops below $T_\mathrm{N}$, as suggested in a kagomé antiferromagnet [49] and a van der Waals magnet [50].

Next, we discuss the field evolution of the phase transitions in the AFM phase of Na$_2$Co$_2$TeO$_6$, inferred from the field dependence of $\kappa_{xx}$ and $\kappa_{xy}$ at 5 K (Fig. 5). In $B \parallel a$, the gradual field-induced increase of $\kappa_{xx}$ and $\kappa_{xy}^a/T$ suggests that the zero-field phase (denoted as phase I shown in Fig. 5(c) and 5(d)) seems to persist up to $B_{\mathrm{c1}}$. The first sign flip of $\kappa_{xy}^a$ at $B_{\mathrm{c1}}$ and the peak in the field dependence of $\kappa_{xx}$ show a magnetic phase transition at $B_{\mathrm{c1}}$ into a different magnetic phase (phase II) with a dominant negative Berry phase distribution and a field-induced suppression on $\kappa_{xx}$. The next sign flip of $\kappa_{xy}^a$ at $B_{\mathrm{c2}}$ along with the marked change in the field dependence of $\kappa_{xx}$ indicates the existence of another magnetic phase (phase III) above $B_{\mathrm{c2}}$. We note that the good correspondence between the field dependence of $\kappa_{xx}$ and $\kappa_{xy}^a/T$ is observed not only at 5 K but also at higher temperatures (see Fig. S9 in SM). These sign changes of the Berry phase distributions between the different magnetic states, suggested by our planar THE measurements, provide essential information for future studies to identify the magnetic states by calculating $\kappa_{xy}$ of them.

In $B \parallel a^*$, the sharp jumps both in $\kappa_{xx}$ and $\kappa_{xy}^{a^*}/T$ indicate the first-order phase transition into phase II* at $B_{\mathrm{c1}}^*$ under $B \parallel a^*$. The sharp jump in $\kappa_{xx}$ is more pronounced in lower temperatures (Fig. 2(b)), whereas the magnitude of $\kappa_{xy}^{a^*}/T$ shows a peak at around 7 K (Fig. 4(a)). This different thermal behavior is caused by the disparate transport processes between $\kappa_{xx}$ and $\kappa_{xy}$. The former is dictated by the suppression of the magnon-phonon scattering by opening a magnon gap, which is more pronounced at lower temperatures. On the other hand, the latter reflects the boson occupation in the energy bands with a finite Berry curvature, showing a peak when the thermal energy coincides with the energy bands with the largest Berry curvature [19]. In contrast to the case in $B \parallel a$, both the field dependences of $\kappa_{xx}$ and $\kappa_{xy}^{a^*}$ in $B \parallel a^*$ are essentially featureless above $B_{c1}^*$, except for small bumps in $\kappa_{xx}$. Given the clear anomalies in $dM/dB$ at $B_{c2}^*$ and $B_{c3}^*$ [33], the featureless field dependence of $\kappa_{xy}^{a^*}/T$ is rather an unexpected result. These results suggest that the phase II* continuously changes to the spin polarized phase, which should be scrutinized by future NMR or neutron scattering measurements. The negative $\kappa_{xy}^{a^*}$ in the phase II* shows the emergence of topological magnon bands with a negative Chern number in this phase II*. Remarkably, both the magnitude and the field dependence of $\kappa_{xy}^{a^*}/T$ are very similar to those of $\kappa_{xy}^{a}/T$ for 8–10 T in the phase II, whereas the field effect on $\kappa_{xx}$ is opposite in these two phases. This similarity in $\kappa_{xy}$ with anisotropic field dependence in $\kappa_{xx}$ may suggest that the phase II and II* share a similar Chern number distribution despite the different magnetic structures between them. Such different magnetic structures in different field directions in the 2D honeycomb plane are shown by the numerical calculations [14,43], and are indeed observed in α–RuCl$_3$ by the neutron scattering experiments under $B \parallel [110]$ and $B \parallel [100]$ [51].

Finally, we discuss the magnetic state above $B_{c3}$ and $B_{c3}^*$. Although the previous studies [30,33] have suggested the emergence of a disordered state above $B_{c3}$ or $B_{c3}^*$, the finite $\kappa_{xy}^{a^*}/T$ above $B_{c1}^*$ reveals the presence of magnetic order that breaks the two-fold rotation symmetry around the $a^*$ axis in $B \parallel a^*$. Given the continuous field dependence of $\kappa_{xy}^{a^*}/T$ and $\kappa_{xx}$ for $B \parallel a^*$, this ordered phase persists above $B_{c1}^*$ up to $B_{sat}$. On the other hand, the sign reversal of $\kappa_{xy}^{a}$ at $B_{c3}$ shows an appearance of a different magnetic state for $B_{c3} < B < B_{sat}$ for $B \parallel a$. The positive $\kappa_{xy}^{a}$ and the vanishing $\kappa_{xy}^{a}/T$ in the zero-temperature limit at 11 T (Fig. 4(a)) are consistent with another magnetically ordered state which possesses topological magnons with a positive Chern number. However, the absence of a clear thermodynamical phase transition above $B_{c3}$ may suggest that the phase III is a disordered state including a partially polarized state [33], requiring further microscopic measurements to reveal the details of the phase III.

In summary, from our planar thermal Hall measurements in Na$_2$Co$_2$TeO$_6$, we find a finite planar $\kappa_{xy}$ with sign changes at critical magnetic fields applied along the $a$ and $a^*$ axes, suggesting a THE of topological magnons. The finite planar $\kappa_{xy}$ in $B \parallel a^*$ reveals that a magnetically ordered state breaking the two-fold rotation symmetry around the $a^*$ axis appears above $B_{c1}^*$, which persists up to $B_{sat}$. We further find three different magnetic phases in $B \parallel a$ which can be characterized by the different Chern numbers of the topological magnons. Our finding, especially for the sign changes of the planar $\kappa_{xy}$, put strong constraints on identifying the magnetic state in this compound.

Figures and legends:

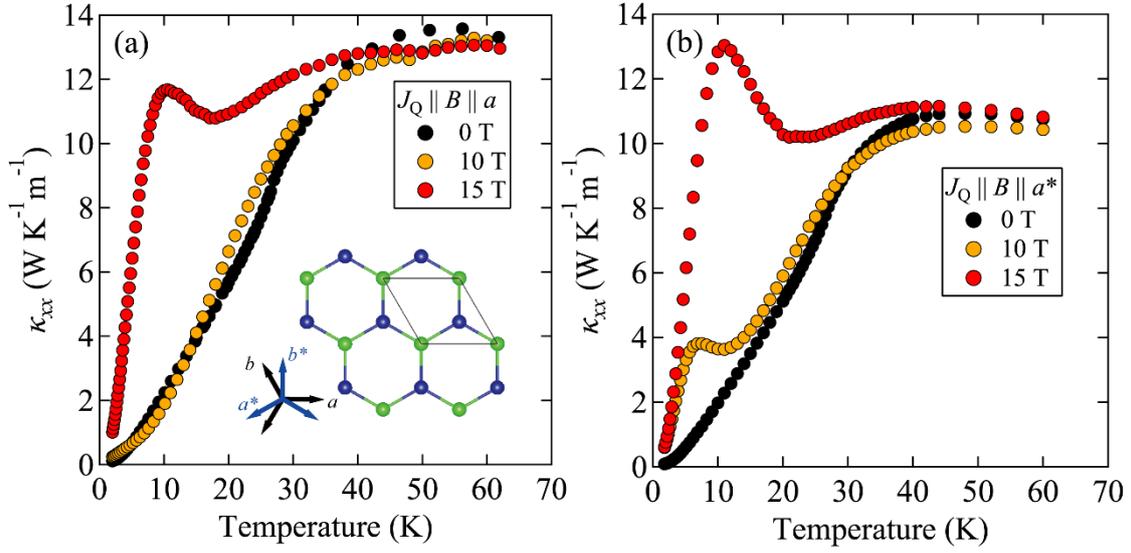

Fig. 1. The temperature dependence of the longitudinal thermal conductivity ($\kappa_{xx}$) under various magnetic fields applied along the $a$ (a) and $a^*$ (b) axes obtained in the VTI measurements. The inset in (a) illustrates the honeycomb lattice of $Co^{2+}$ ions with the crystallographic axes.

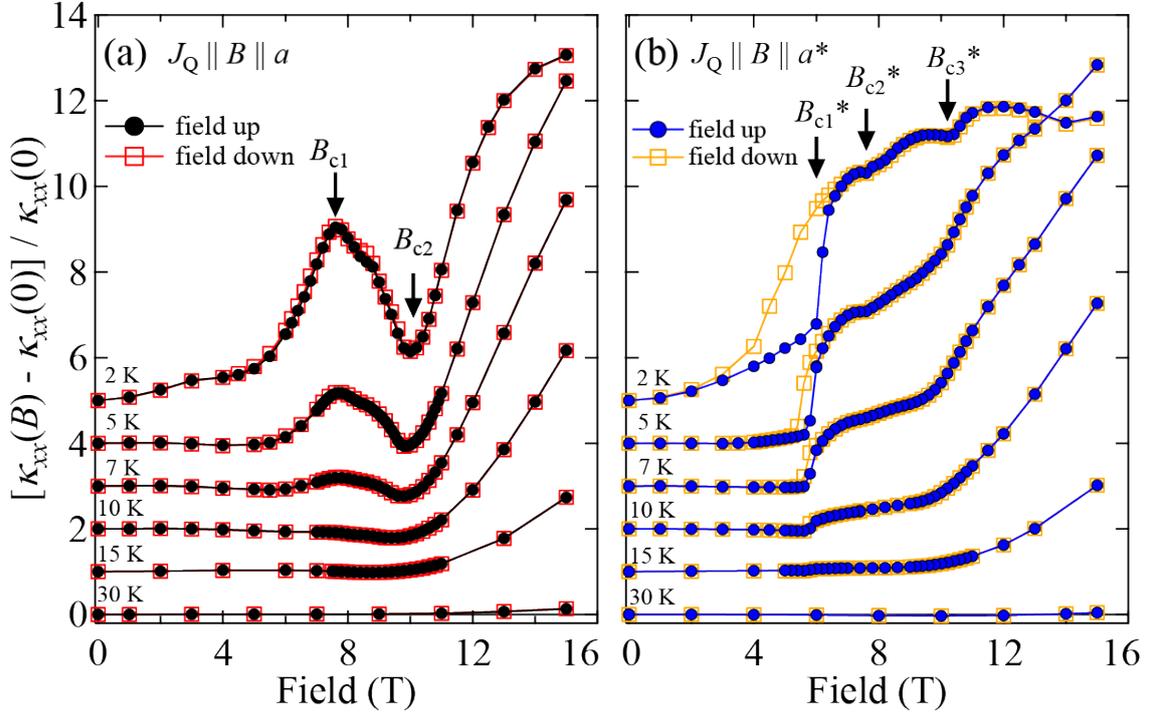

Fig. 2. The field-induced deviation of the longitudinal thermal conductivity $[\kappa_{xx}(B) - \kappa_{xx}(0)]$ normalized by the zero-field value under magnetic fields applied along the $a$ (a) and $a^*$ (b) axes. For clarity, the vertical axis of the data $T \leq 15$ K is shifted. The data obtained in the field-up and field-down procedure is shown in filled and open symbols, respectively.

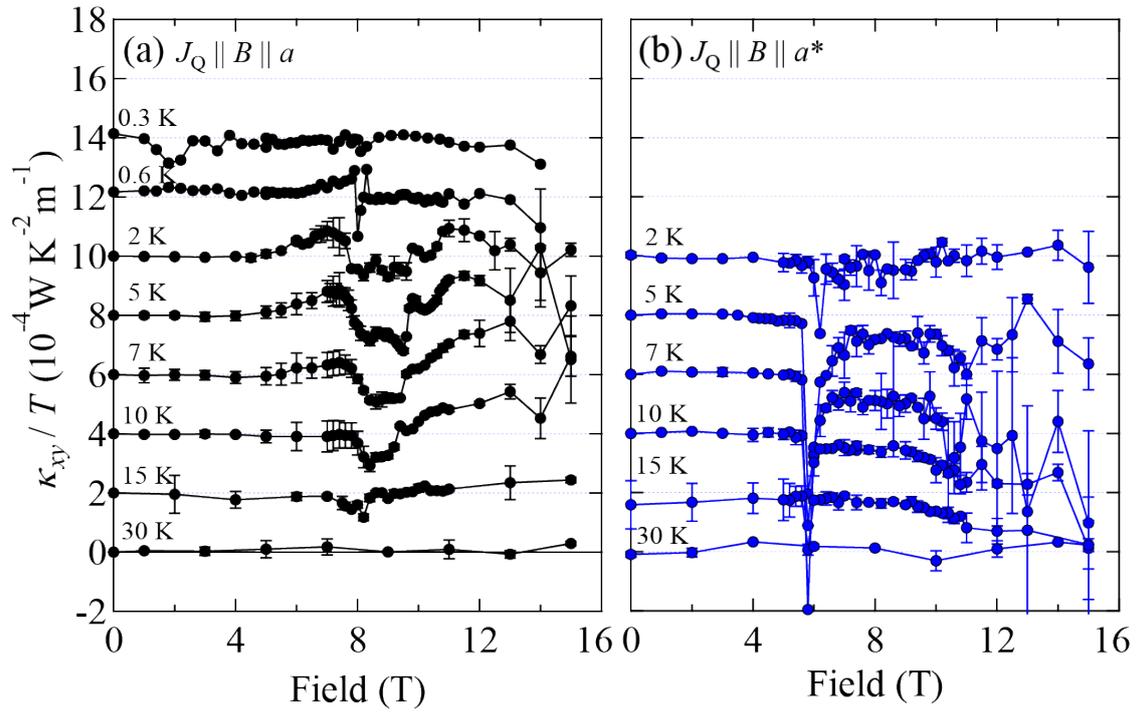

Fig. 3. The field dependence of the thermal Hall conductivity divided by the temperature for $B \parallel a$ (a) and $B \parallel a^*$ (b) at different temperatures. The data is shifted by a constant amount for clarity. The error bars show the deviation of the data in the field-up and field-down measurements.

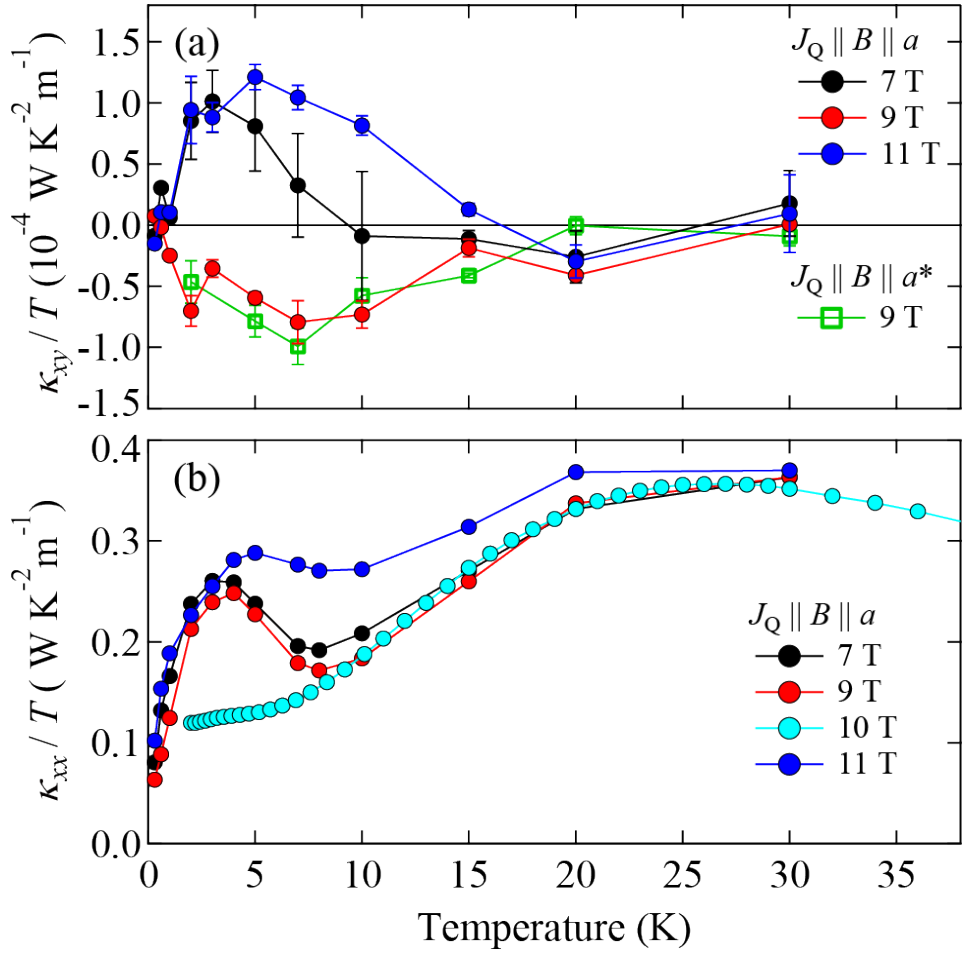

Fig. 4. The temperature dependence of $\kappa_{xy}/T$ (a) and $\kappa_{xx}/T$ (b) at 7, 9, 11 T for $B \parallel a$, and at 9 T for $B \parallel a^*$. The temperature dependence of $\kappa_{xx}/T$ at 10 T, the same data shown in Fig. 1(a), is also shown for comparison.

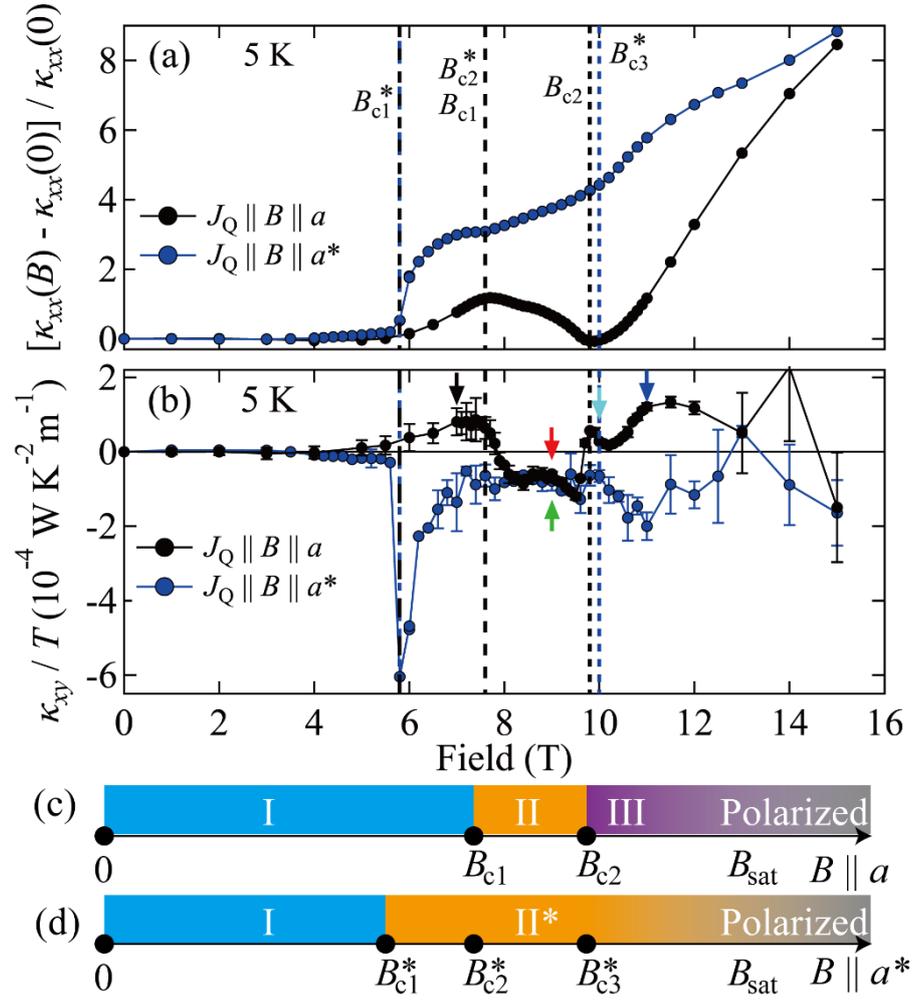

Fig. 5. (a,b) The comparison of the field dependence of the longitudinal thermal conductivity ($\kappa_{xx}$, a) and the planar thermal Hall conductivity ($\kappa_{xy}$, b) for $B \parallel a$ (black) and $B \parallel a^*$ (blue) at 5 K. The critical fields are also marked by dotted lines. The arrows indicate the field used in the data of the same color in Fig. 4. (c,d) The suggested magnetic phases of $Na_2Co_2TeO_6$ in $B \parallel a$ (c) and $B \parallel a^*$ (d), drawn from our thermal-transport measurements.

Acknowledgements:

We thank J. Nasu for fruitful discussions. This work was supported by Grants-in-Aid for Scientific Research (KAKENHI) (No. JP19H01848 and No. JP19K21842) and the Murata Science Foundation. M.A. was supported by JST SPRING, Grant Number JPMJSP2108. R.S. acknowledges the financial support provided by the Ministry of Science and Technology in Taiwan under Project No. MOST-110-2112-M-001-065-MY3 and Academia Sinica budget of AS-iMATE-111-12. The work at SKKU is supported by the National Research Foundation (NRF) of Korea (Grant no. 2020R1A2C3012367 and 2020R1A5A1016518).

# Supplementary Materials for "Planar Thermal Hall Effects in Kitaev Spin Liquid Candidate Na$_2$Co$_2$TeO$_6$"

Hikaru Takeda, Jiancong Mai, Masatoshi Akazawa, Kyo Tamura, Jian Yan, Kalimuthu Moovendaran, Kalaivanan Raju, Raman Sankar, Kwang-Yong Choi, and Minoru Yamashita

## I. The details of the thermal-transport measurements

Figure S1 shows the setup of our thermal-transport measurements. As shown in Fig. 1, one heater and three thermometers ($T_\text{High}$, $T_\text{L1}$, and $T_\text{L2}$) were attached to the sample with a silver paste. The sample size is about $2.1 \times 1.7 \times 0.12$ mm$^3$. To avoid a background signal coming from a metal, the sample was attached to the insulating LiF heat bath with non-metallic grease. The heat current $J_\text{Q}$ and the magnetic field $B$ were applied parallel to either the $a$ and $a^*$ axes of the sample. Note that the positive $B$ direction was defined as the direction of $J_\text{Q}$ (i.e. $-x$ direction, see Fig. S1(a)). Both the longitudinal $\Delta T_x$ ($\Delta T_x = T_\text{H} - T_\text{L2}$) and the transverse $\Delta T_y$ ($\Delta T_y = T_\text{L1} - T_\text{L2}$) temperature difference were measured as a function of the heat current $J_\text{Q} = Q/wt$. To cancel the longitudinal component in $\Delta T_y$ by the misalignment effect, $\Delta T_y$ was asymmetrized with respect to the field direction as $\Delta T_y^{asym} = \Delta T_y(+B) - \Delta T_y(-B)$. To take into account a magnetic hysteresis effect, this antisymmetrization was done separately for the data obtained in the field-up process and that in the field-down process. The DR measurements were done only in the field-down process in $B \parallel J_\text{Q} \parallel a$. The thermal (Hall) conductivity $\kappa_{xx}$ ($\kappa_{xy}$) is derived by

$$\begin{pmatrix} Q/wt \\ 0 \end{pmatrix} = \begin{pmatrix} \kappa_{xx} & \kappa_{xy} \\ -\kappa_{xy} & \kappa_{xx} \end{pmatrix} \begin{pmatrix} \Delta T_x/L \\ \Delta T_y^{asym}/w' \end{pmatrix},$$

where $t$ is the thickness of the sample, $L$ is the length between $T_\text{H}$ and $T_\text{L1}$, and $w$ is the sample width between $T_\text{L1}$ and $T_\text{L2}$, and $w'$ is the length between $T_\text{L1}$ and $T_\text{L2}$.

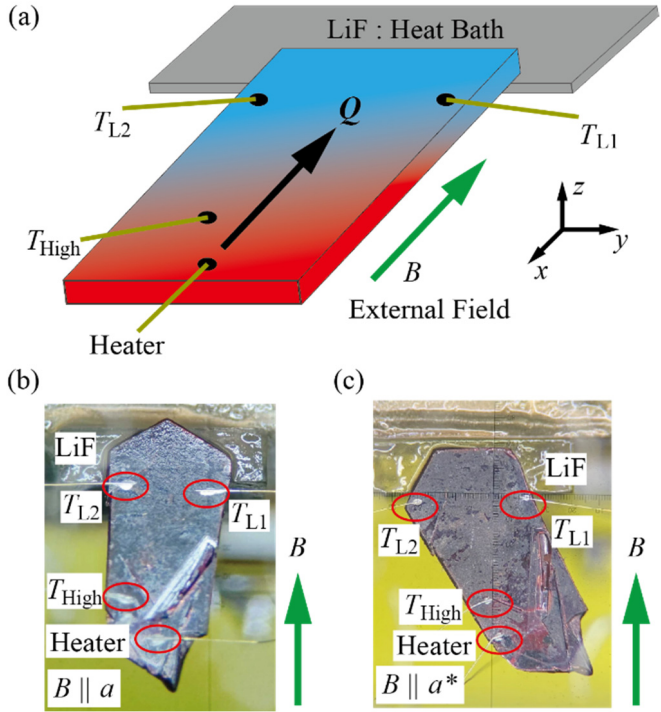

Fig. S1 (a) A schematic illustration of the experimental setup. (b, c) Photographs showing the setup for the measurement of $J_Q \parallel B \parallel a$ (b) and $J_Q \parallel B \parallel a^*$ (c).

## II. Extended data of the magnetization and the thermal conductivity

Figure S2 shows the temperature dependence of $\kappa_{xx}/T$ of sample #1 (the same data of Fig. 1 in the main text). As shown in Fig. S2, the kink in the temperature dependence of $\kappa_{xx}/T$ is clearly seen at the Néel temperature ($T_N$) at 0 T.

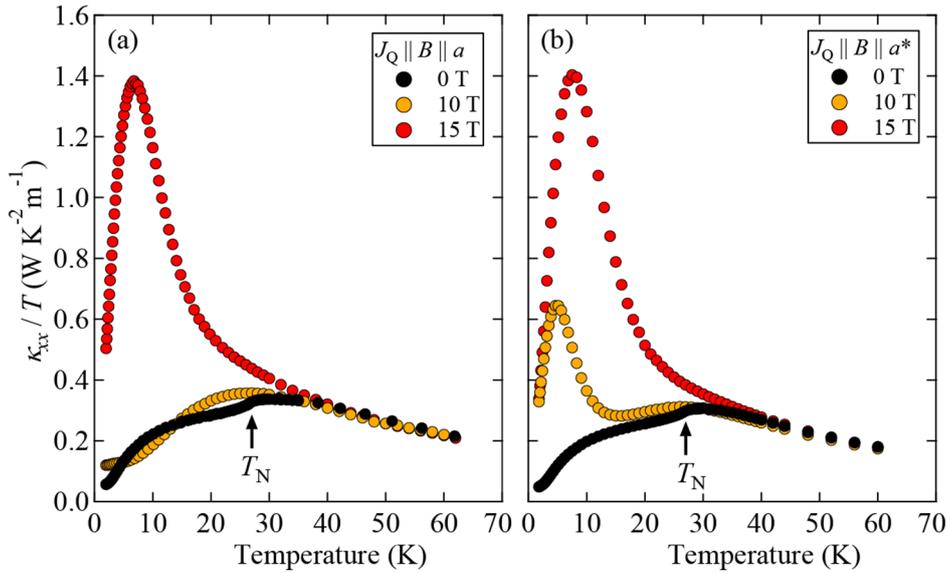

Fig. S2 (a,b) The temperature dependence of $\kappa_{xx}/T$ of sample #1 at 0, 10, and 15 T for $J_Q \parallel B \parallel a$ (a) and $J_Q \parallel B \parallel a^*$ (b).

Temperature dependence of the magnetic susceptibility of sample #1 was measured at 1 T for both $B \parallel a$ and $B \parallel a^*$ (Fig. S3(a)). As shown in Fig. S3(a), the Néel transition is clearly observed as the peak. The field dependence of the magnetization was measured at 2 K (Fig. S3(b)). The sharp jump in the magnetization is observed at $B_{c1}^*$ in $B \parallel a^*$. In $B \parallel a$, only the gradual increase is observed up to 7 T. These magnetization data well reproduce the previous results [1–3], showing the good quality of our sample.

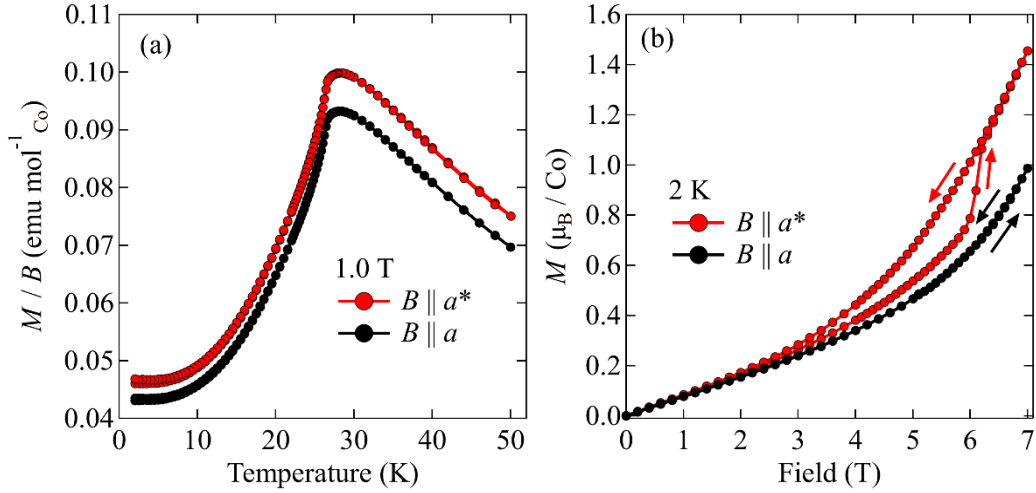

Fig. S3 (a) The temperature dependence of the magnetic susceptibility of sample #1 at 1.0 T applied along the $a$ (black) and $a^*$ (red) axes. (b) The field dependence of the magnetization ($M$) at 2 K.

### III. The sample dependence of the thermal Hall effect

In order to check sample dependence of the thermal Hall effect, we measured $\kappa_{xx}$ and $\kappa_{xy}$ of another sample (sample #2) for both $J_Q \parallel B \parallel a$ and $J_Q \parallel B \parallel a^*$, as well as the field dependence of $M$ in $B \parallel a$ and $B \parallel a^*$. Here we compare some experimental results of sample #2 with those of sample #1 shown in the main text.

Figure S4 shows the temperature dependence of $\kappa_{xx}$ at $B = 0$ (Fig. S4a) and the magnetic field dependence of $M$ at 2 K (Fig. S4b). As shown in Fig. S4, both $\kappa_{xx}(T)$ and $M(B)$ of sample #2 are similar to those of sample #1, showing a good reproducibility of our data. The slight reduction of the magnitude of $\kappa_{xx}$ suggests a shorter mean free path of phonons owing to a lower sample quality of sample #2.

Figure S5 compares the field dependence of $\kappa_{xy}/T$ of sample #1 and that of sample #2 at 5 and 7 K for both $B \parallel a$ (Fig. S5a) and $B \parallel a^*$ (Fig. S5b). Although the absolute values for sample #2 are smaller than those for sample #1 due to the lower sample quality, both samples show almost the same magnetic field dependence, showing a good reproducibility of our data against the sample quality. This good reproducibility demonstrates that the features observed in the field dependence of $\kappa_{xy}$, such as the sign reversals of $\kappa_{xy}$ for $B \parallel a$ and the presence of the finite $\kappa_{xy}$ for $B \parallel a^*$, are intrinsic properties of this material. We stress that the smaller $\kappa_{xy}$ of sample #2 is consistent with the absence of a fermionic contribution in $\kappa_{xy}$. Both results that lead us to conclude the absence – (1) $\kappa_{xy}$ is one order of magnitude smaller than the half-quantized value expected for the Kitaev excitations (2) $\kappa_{xy}/T$ asymptotically approaches zero as lowering temperature – are well reproduced in sample #2.

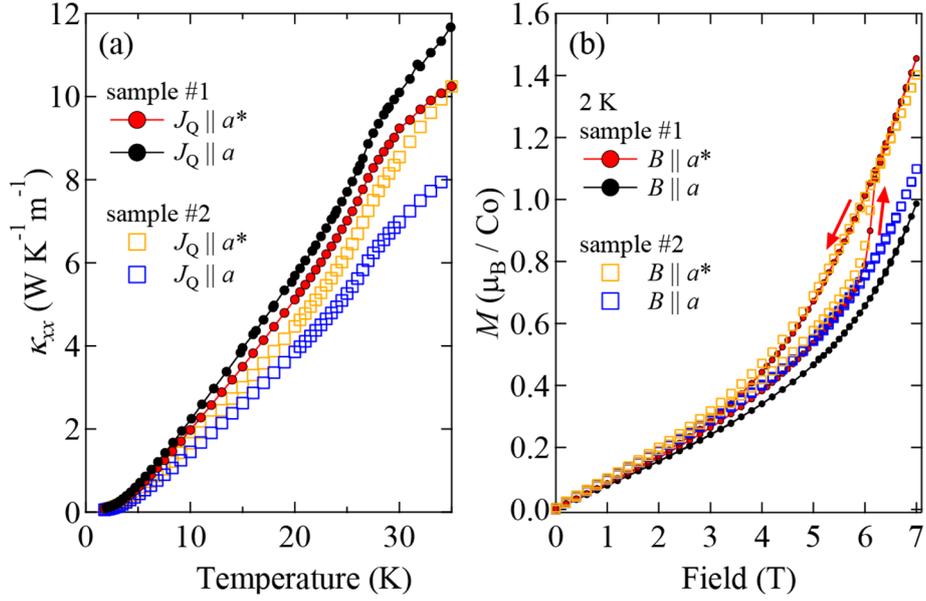

Fig. S4 (a) The temperature dependence of the longitudinal thermal conductivity of sample #1 and sample #2 at 0 T. (b) The field dependence of the magnetization of sample #1 and sample #2 at 2 K. The $\kappa_{xx}$ ($M$) data of sample #1 is the same as that shown in Fig. 1 in the main text (Fig. S3(b)).

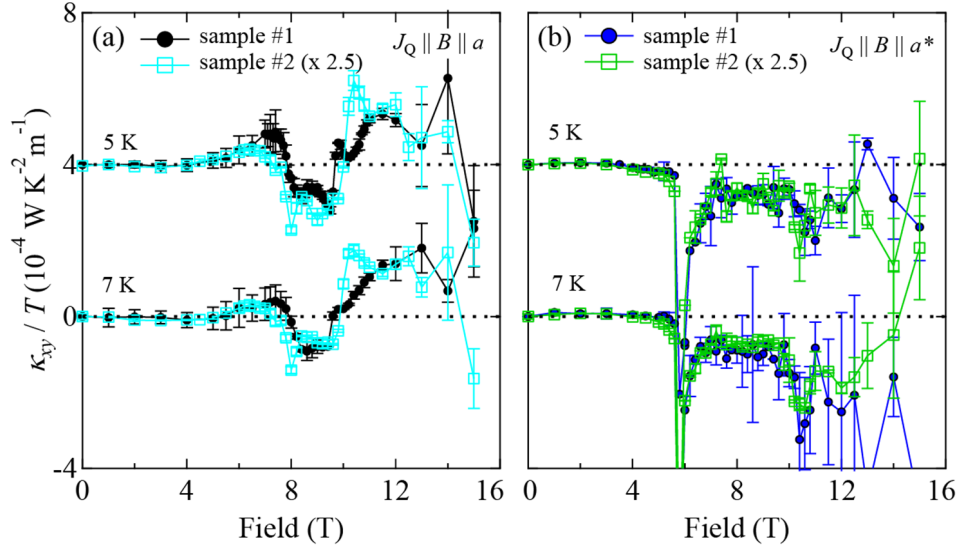

Fig. S5 (a,b) The field dependence of $\kappa_{xy}/T$ at 5 and 7 K of sample #1 and sample #2 measured in $J_\mathrm{Q} \parallel B \parallel a$ (a) and $J_\mathrm{Q} \parallel B \parallel a^*$ (b). The data of sample #2 is multiplied by 2.5 for clarity.

## IV. Estimation of the phonon mean free path

As shown in Fig. 1 in the main text and Fig. S2, the in-plane magnetic field increases $\kappa_{xx}$ for both field directions. The temperature dependence of $\kappa_{xx}$ at 15 T shows a double-peak structure with the lower temperature peak around 10 K, which is given by the field-enhancing effect on phonons by suppressing the magnon-phonon scattering by the magnetic field [3]. The exponent of the temperature dependence below 0.4 K is increased from 1.8 at 0 T to 2.4 under 14 T (Fig. S6(a)), which is close to that of boundary-limited phonons ($\kappa_{xx} \propto T^3$). In fact, the long mean free path of phonons is obtained as following. The thermal conductivity of phonons is given by

$$\kappa_{xx}^{\mathrm{ph}} = \frac{1}{3} C_{\mathrm{ph}}\, v_{\mathrm{ph}}\, \ell_{\mathrm{ph}},$$

where $C_{\mathrm{ph}}$, $v_{\mathrm{ph}}$, and $\ell_{\mathrm{ph}}$ is the heat capacity, the sound velocity, and the mean free path of phonons, respectively. We estimate $C_{\mathrm{ph}}$ by a $T^3$ fitting to the heat capacity data of the nonmagnetic Na$_2$Zn$_2$TeO$_6$ [1] as $\beta = C_{\mathrm{ph}}/T^3 = 4.8 \times 10^{-4}$ J K$^{-4}$ mol$^{-1}$. The sound velocity is estimated as approximately 3500 m/s from the value of $\beta$. Given the dominant phonon contribution in $\kappa_{xx}$ [3], the temperature dependence of the phonon mean free path (Fig. S6(b)) is obtained from $\kappa_{xx}$ (Fig. S6(a)). As shown in Fig. S6(b), the phonon mean path exceeds the thickness of the sample (approximately 0.1 mm) below 0.2 and 0.5 K at 0 and 14 T, respectively. The temperature dependence of $\ell_{\mathrm{ph}}$ at 14 T seems to saturate near the temperature-independent mean free path $\ell_0 = 2\sqrt{A/\pi} \sim$ 0.5 mm at the lowest temperature, where $A$ is the cross-sectional area of the sample. This long phonon mean free path demonstrates the good quality of the sample.

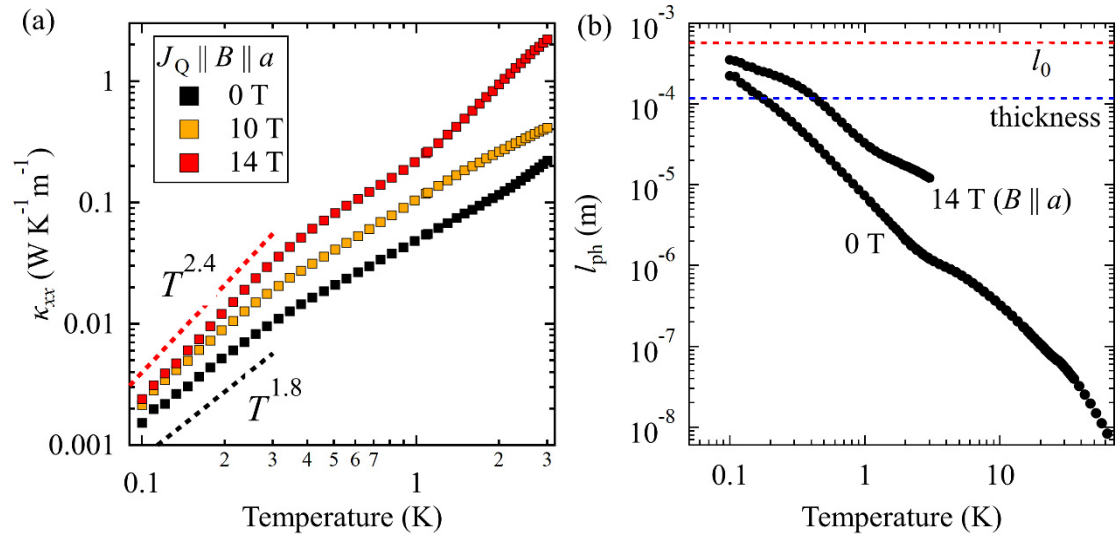

Fig. S6 (a) A log-log plot of the temperature dependence of $\kappa_{xx}$ of sample #1 measured in the DR measurements. (b) The temperature dependence of the phonon mean free path at 0 and 14 T for $B \parallel a$.

## V. The residual of $\kappa_{xx}/T$ in the zero-temperature limit

Figure S7 shows the temperature dependence of $\kappa_{xx}/T$ at different magnetic fields obtained in the DR measurements. As shown in Fig. S7, all the extrapolation of $\kappa_{xx}/T$ becomes negligibly small at $T = 0$ K, showing the absence of itinerant gapless excitations in Na$_2$Co$_2$TeO$_6$.

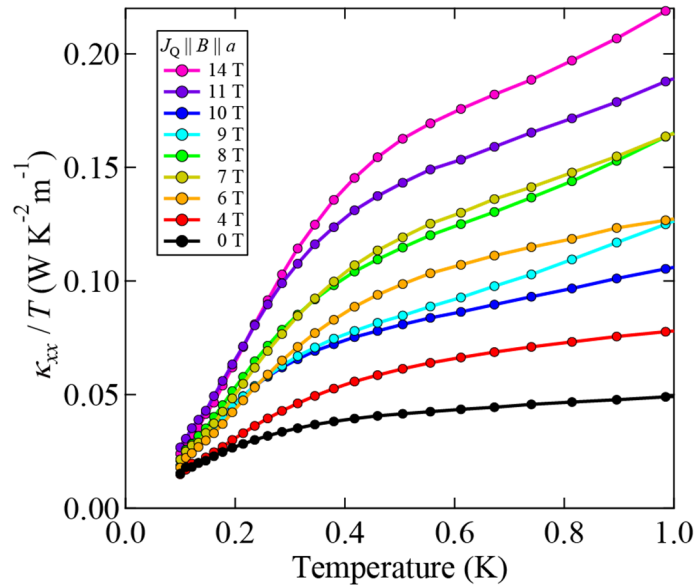

Fig. S7 The temperature dependence of $\kappa_{xx}/T$ of sample #1 at different magnetic fields applied along the $a$ axis.

## VI. Field effect on the phonon contribution and the magnon contribution in $\kappa_{xx}$

As $\kappa_{xx}$ of a magnetic insulator is given by the sum of the phonon contribution $\kappa_{xx}^{\mathrm{ph}}$ and the magnon contribution $\kappa_{xx}^{\mathrm{mag}}$, the field dependence of $\kappa_{xx}$ of this compound can be understood in terms of the field effect of the magnon gap on $\kappa_{xx}^{\mathrm{ph}}$ and $\kappa_{xx}^{\mathrm{mag}}$. When the magnon gap is increased by applying a magnetic field (e.g. in a forced-ferromagnetic phase), $\kappa_{xx}^{\mathrm{ph}}$ is increased by the suppression of magnon-phonon scatterings and $\kappa_{xx}^{\mathrm{mag}}$ is decreased by the reduction of the populations of excited magnons, and vice versa. Therefore, the field-induced increases of $\kappa_{xx}$ observed both above $\sim 6$ T in $B \parallel a$ and $B_{c1}^*$ in $B \parallel a^*$ are well understood by the increase of $\kappa_{xx}^{\mathrm{ph}}$ due to an increase of the magnon gap (Fig. 2 in the main text). In this compound, it is suggested that an AFM zigzag ordered state with collinear moments pointing to the zigzag chain ($\parallel a$) is supplemented with a minor staggered moment along the $a^*$ direction [1]. The sudden increase of the magnetization at $B_{c1}^*$ in $B \parallel a^*$ (Fig. S3(b)) is explained by the spin-flop transition of the minor staggered moment. This sudden increase of the magnetization also increases the magnon gap induced by the spin anisotropy, resulting in the sudden increase of $\kappa_{xx}^{\mathrm{ph}}$ by the decrease magnon-phonon scatterings. This magnon gap may gradually increase in $B \parallel a$ as the magnetization does, consistent with the gradual increase of $\kappa_{xx}^{\mathrm{ph}}$ in $B \parallel a$.

For $B \parallel a^*$, the field increase effect on $\kappa_{xx}$ above $B_{c1}^*$ continues up to the highest measuring field (Fig. 2(b) in the main text), showing the field-induced gap emerging at $B_{c1}^*$ persists to $B_{\mathrm{sat}}$. On the other hand, $\kappa_{xx}$ for $B \parallel a$ decreases above $B_{c1}$ until it again turns to increase above $B_{c2}$ (Fig. 2(a) in the main text). This minimum of $\kappa_{xx}$ at $B_{c2}$ suggests that the magnon gap once closes at $B_{c2}$ at the end of the AFM phase, which is followed by opening of a field-induced gap in the disordered state above $B_{c2}$. This contrasting field dependence of $\kappa_{xx}$ in $B \parallel a$ and that in $B \parallel a^*$ demonstrates a different magnetic state in $B \parallel a$ from that above $B_{c1}^*$ in $B \parallel a^*$.

## VII. The field dependence of $\kappa_{xy}/T$ in the field-up and the field-down measurements

Figure S8 shows the field dependence of $\kappa_{xy}/T$ in $B \parallel a$ and in $B \parallel a^*$ measured in the field-up (filled symbols) and the field-down (open symbols) measurements. As shown in Fig. S8, both the data show essentially the same field dependence, except for the clear hysteresis in $B \parallel a^*$ at $B_{c1}^*$ caused by the first-order transition.

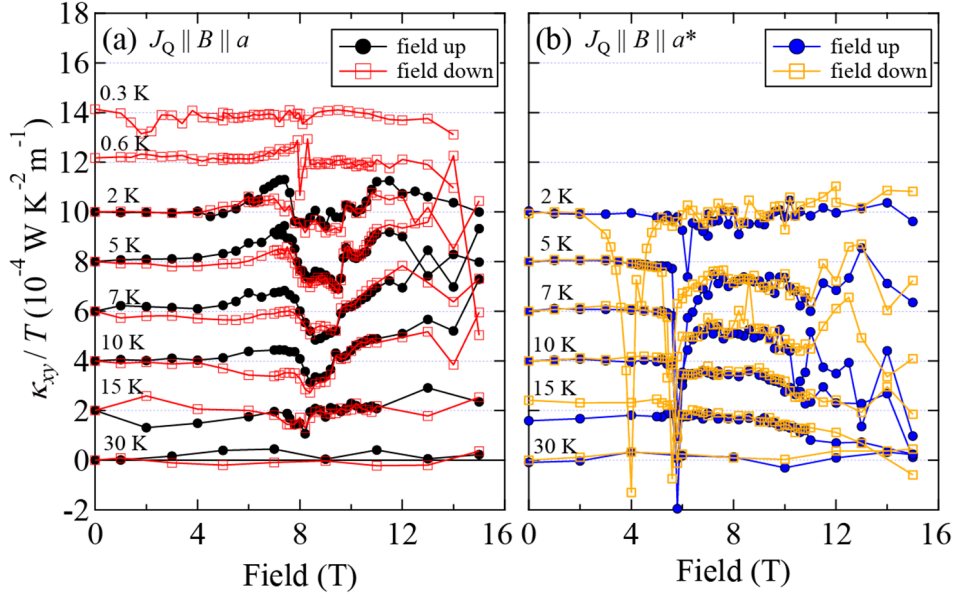

Fig. S8 The field dependence of $\kappa_{xy}/T$ of sample #1 obtained in the field-up (filled circles) and the field-down (open squares) measurements in $J_Q \parallel B \parallel a$ (a) and $J_Q \parallel B \parallel a^*$ (b). The data in the DR measurements ($T < 2$ K) was obtained only in the field-down process in $J_Q \parallel B \parallel a$.

## VIII. The field dependence of $\kappa_{xx}$ and $\kappa_{xy}/T$ at high temperatures

Figure S9 shows an enlarged view of the field dependence of $\kappa_{xx}$ and $\kappa_{xy}/T$ at high temperatures. Although the peak in $\kappa_{xx}(B)$ at $B_{c1}$ is suppressed at 10 and 15 K as shown in Fig. S9(a), the minimum at $B_{c2}$ is clearly observed at a smaller $B$ at a higher $T$. As shown in Fig. S9(c), the anomalies in $\kappa_{xx}(B)$ have a good correspondence to the changes observed in $\kappa_{xy}/T$.

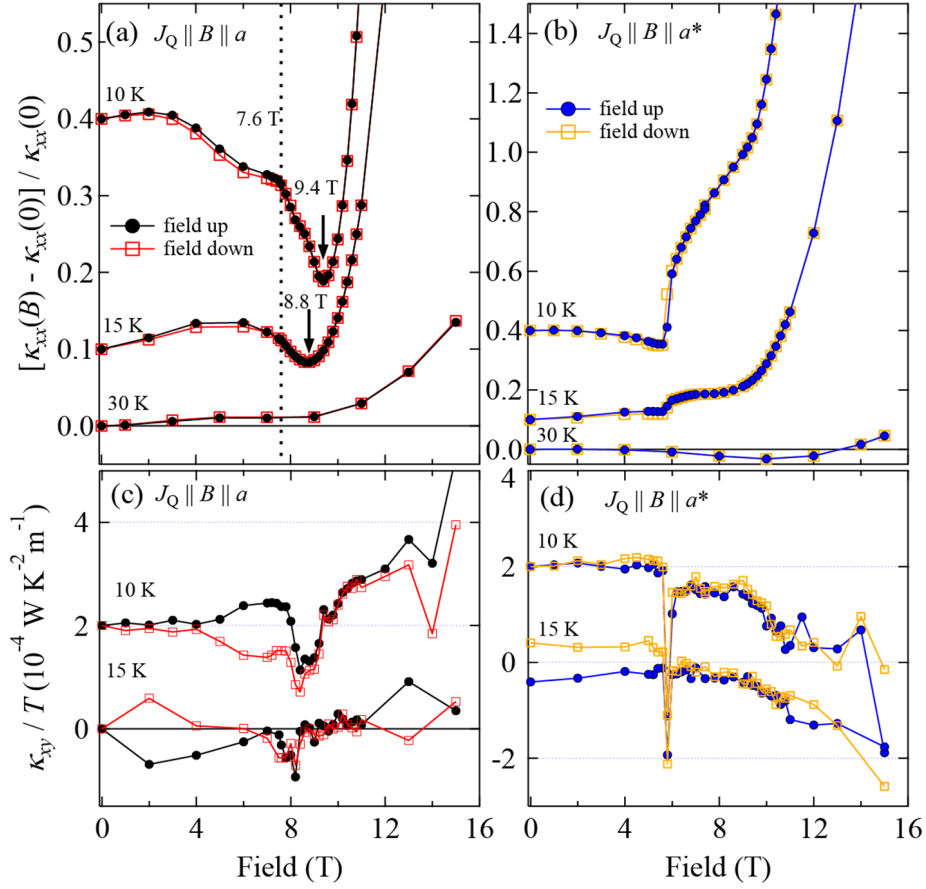

Fig. S9 An enlarged view of Fig. 2 and Fig. 3 in the main text for the high temperature data.

## IX. Supplementary References

*at Small Hall Angles*, Phys. Rev. Lett. **121**, 147201 (2018).

[8] R. Hentrich et al., *High-Field Thermal Transport Properties of the Kitaev Quantum Magnet α–RuCl$_3$: Evidence for Low-Energy Excitations beyond the Critical Field*, Phys. Rev. B **102**, 235155 (2020).